\documentclass[graybox]{svmult}

\usepackage{mathptmx}       % selects Times Roman as basic font   maha se za po-krasivo ...
\usepackage{helvet}         % selects Helvetica as sans-serif font
\usepackage{courier}        % selects Courier as typewriter font
\usepackage{type1cm}        % activate if the above 3 fonts are
\usepackage{makeidx}         % allows index generation
\usepackage{graphicx}        % standard LaTeX graphics tool
\usepackage{multicol}        % used for the two-column index
\usepackage[bottom]{footmisc}% places footnotes at page bottom
\usepackage{amsmath}
\begin{document}

\title*{The Heun functions and their applications in astrophysics.}
% Use \titlerunning{Short Title} for an abbreviated version of
% your contribution title if the original one is too long
\author{Denitsa Staicova and Plamen Fiziev}
% Use \authorrunning{Short Title} for an abbreviated version of
% your contribution title if the original one is too long
\institute{Institute for Nuclear Research and Nuclear Energy,
\at Bulgarian Academy of Sciences, bul. ``Tsarigrasko shose'' 72, Sofia 1784, Bulgaria, \email{dstaicova@inrne.bas.bg}
\and Sofia University Foundation for Theoretical and Computational Physics and
Astrophysics, \at 5 James Bourchier Blvd., 1164 Sofia, Bulgaria and
JINR, Dubna, 141980 Moscow Region, Russia \email{fiziev@phys.uni-sofia.bg}}
\maketitle

\abstract{The Heun functions are often called the hypergemeotry successors of the 21st century, because of the wide number of their applications. In this proceeding we discuss their application to the problem of perturbations of rotating and non-rotating black holes and highlight some recent results on their late-time ring-down obtained using those functions.}

\section{The Heun Functions}
\label{sec:1}

The Heun functions are gaining popularity due to the vast number of their applications. The Heun project, a site dedicated to gathering scientists working in this area, has already accumulated more than 500 articles on the theory and the applications of those functions. Among the topics are the Schr{\"o}dinger equation with anharmoic potential,  the Teukolsky linear perturbation theory for the Schwarzschild and Kerr metrics, transversable wormholes, quantum Rabi models, confinement of graphene electrons in different potentials, quantum critical systems, crystalline materials, three-dimensional atmospheric and ocean waves, single polymer dynamics,  economics, genetics e.t.c (see the bibliography section in \cite{site}). 

The general Heun function is defined as the local solution of the following second order Fuchsian ordinary differential equation (ODE) \cite{HC1,HC2}:

\begin{equation}
\frac{d^2}{dz^2}H(z)+\left[\frac{\gamma}{z}+\frac{\delta}{z-1}+\frac{\epsilon}{z-a}\right]\frac{dH(z)}{dz}+\frac{\alpha \beta z-q}{z(z-1)(z-a)}H(z)=0
\normalsize
\label{eq1}
\end{equation}
normalized to 1 at $z=0$ % and regular at $z=0$. 
Here  $\epsilon=\alpha+\beta-\gamma-\delta+1$. This equation posses 4 regular singularities: $z=0, 1, a, \infty$ and it generalizes the hypergeometric function, the Lam{\'e} function, the Mathieu function, the spheroidal wave functions etc. Its group of symmetries is of order 192.

For comparison, the hypregeometric differential equation has 3 regular singularities $z(z-1)\frac{d^2 w(z)}{dz^2}+\left[c-(a+b+1)z\right]\frac{d w(z)}{dz}-abw(z)=0$ with group of symmetries  of order 24.

Recalling the definition of irregular singularity: 
\begin{definition}
For an ODE of the form: $P(x)y''(x)+Q(x)y'(x)+R(x)y(x)=0$, the point $x_0$ is singular if Q(x)/P(x) or R(x)/P(x) diverge at $x=x_0$. If the limits $\lim_{x\to x_0}\frac{Q(x)}{P(x)}(x-x_0)$ and  $\lim_{x\to x_0}\frac{R(x)}{P(x)}(x-x_0)^2$ exist and are finite then the point $x_0$ is regular singularity, otherwise, it is irregular or essential singularity. The point $x_0=\infty$ is treated the same way under the change $x=1/z$.

\end{definition}

\begin{figure}[h]
\sidecaption
\includegraphics[scale=.75]{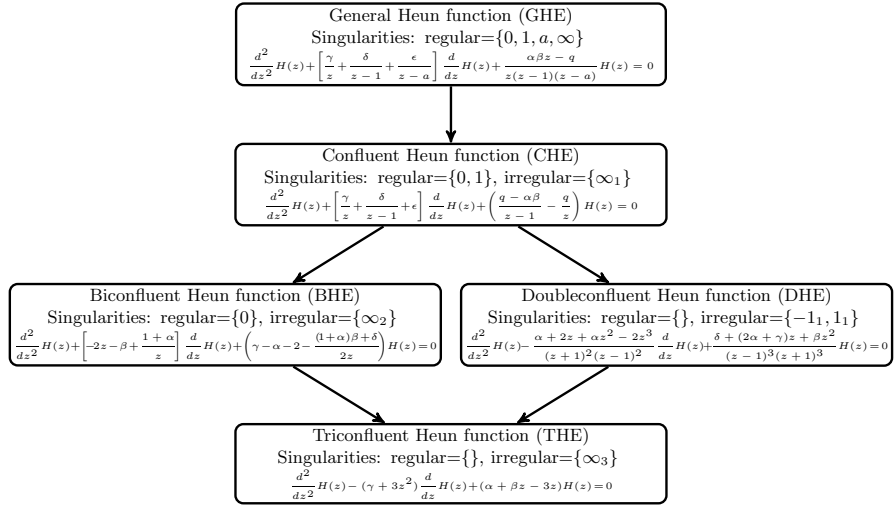}
\caption{A scheme of the different confluent ODEs obtainable from the ODE of the general Heun function (in Maple's notations). The subscript next to the irregular singularities is their rank}
\label{fig:1}       % Give a unique label
\end{figure}

The general Heun function has 4 regular singularities, from which under the process called confluence of singularities, one obtains 4 different types of confluent Heun functions with fewer singularities but of higher s-rank (See Fig. \ref{fig:1} for illustration). 

For the confluent Heun function which we will use below, this process means the redefinition of $\beta=\beta a, \epsilon=\epsilon a, q=qa$ and taking the limit $a\to \infty$. This gives us the following ODE:

\begin{equation}
{\frac {{d}^{2}}{{d}{z}^{2}}}H(z) - \left( \epsilon-{\frac {\delta}{z-1}}-{\frac {\gamma}{z}} \right) {\frac {\rm d}{{\rm d}z}}H(z) - \left( {\frac {\alpha\,\beta-qz}{z-1}}+{\frac {q}{z}} \right) H (z)=0
\label{eq2}
\normalsize
\end{equation}

In Maple notations, the default form of the solution of this type of ODE is denoted as $HeunC(\alpha,\beta,\gamma,\delta,\eta,z)$ which we adopt. To obtain from Maple's default form Eq. \ref{eq1}, one needs to set $\alpha=-(\epsilon_0^2-4q_0)^{1/2}, \beta = \gamma_0-1, \gamma = -1+\delta_0, \delta = -\alpha_0\beta_0+(1/2)\delta_0\epsilon_0+(1/2)\epsilon_0\gamma_0, \eta = -(1/2)\delta_0\gamma_0-(1/2)\epsilon_0\gamma_0+q_0+1/2$ and vise versa (the ``$_0$'' subscript denotes the parameters in Eq. \ref{eq2}.

\section{Applications of the Heun functions in astrophysics}

\subsection{Teukolsky Angular equation and Teukolsky Radial equation }

In the frame of the Teukolsky linear perturbations theory, the late-time ringing of a black hole due to a perturbation of different spin is described by one Master equation. Under the substitution $\Psi(t,r,\theta, \phi)=e^{i(\omega t + m\phi)}S(\theta)R(r)$ (where where $m=0, \pm 1, \pm 2$) this equation splits in two second order ODEs of the confluent Heun type -- The Teukolsky Angular Equation (TAE):
\begin{footnotesize}
 \begin{equation}
 \frac{d}{du}\Big(\left(1\!-\!u^2\right)\frac{d}{du}S_{lm}(u)\Big)+
\left((a\omega u)^2+2a\omega
su\!+\!E\!-\!s^2-\frac{(m\!+\!su)^2}{1\!-\!u^2}\right)S_{lm}(u)=0,
\label{TAE}
\end{equation}
\end{footnotesize}
and the Teukolsky Radial Equation (TRE):
\begin{footnotesize}
\begin{multline}
{\frac {d^{2}R_{l,m}(r)}{d{r}^{2}}} + (1+s)  \left( {\frac{1}{r-{\it r_{+}}}}+
{\frac{1}{r-{\it r_{-}}}} \right){\frac {dR_{l,m}(r)}{dr}} +                                      
+\Biggl( {\frac { K ^{2}}{ \left( r-r_{+} \right) \left( r-r_{-} \right) }}- \\
is \left( {\frac{1}{r-{\it r_{+}}}}+ {\frac{1}{r-{\it r_{-}}}} \right)  K 
-\lambda - 4\,i s \omega r \Biggr)
{\frac {R_{l,m}(r)} {( r-r_{+})( r-r_{-})}}=0
\label{TRE}
\end{multline}
\end{footnotesize}

\noindent \noindent where $\Delta=r^2-2Mr+a^2=(r-r{_-})(r-r{_+})$, $K=-\omega(r^2+a^2)-ma$,
$\lambda=E-s(s+1)+a^2\omega^2+2am\omega$ and $u=\cos(\theta)$. Here $r_{\pm}=M\pm \sqrt{M^2-a^2}$ are the inner and outer horizon of the rotating black hole. Being interested in electromagnetic perturbations we fix the spin to $s=-1$. 

In this system, the unknown quantities are the complex frequencies $\omega_{l,m,n}$ giving us the spectrum and the constant of separation $E_{l,m,n}$ which for $a=0$ is $E=l(l+1)$ (for $s=-1$). The only physical parameters of the system, in agreement with the No-Hair Theorem, are the rotational parameter $a$ and the mass of the black hole $M$, which we here fix to $M=1/2$. 

The singularities of the two equations are as follows: for the TRE $r=r_{\pm}$ -- regular and $r=\infty$ -- irregular. For the TAE, the regular singularities are: $\theta=\pm \pi$ and the irregular is again $\theta=\infty$.

\subsection{Boundary conditions}
In order to find the spectrum, we need to solve the central two-point connection problem, imposing appropriate boundary conditions on two of the singular points. Details on the boundary conditions, as well as on the whole approach and the explicit values of the parameters, can be found in \cite{F1, F2, F3, DF1,DF2,DF3, DF4}. In brief, we require:

\begin{enumerate}
\item{On the TAE:} 
\begin{enumerate}
 \item {Quasi-normal modes (QNMs):} we require  angular regularity. This translate into the following determinant.
\begin{footnotesize}
 \begin{align}
 W[S_1,S_2]=\frac{\text{HeunC}'(\alpha_1,\beta_1,\gamma_1,\delta_1,\eta_1,\left( \cos \left( \pi/6  \right)  \right) ^{2})}{\text{HeunC}(\alpha_1,\beta_1,\gamma_1,\delta_1,\eta_1,\left( \cos \left( \pi/6  \right)  \right) ^{2})}+\notag\\
\frac{\text{HeunC}'(\alpha_2,\beta_2,\gamma_2,\delta_2,\eta_2,\left( \sin \left( \pi/6  \right)  \right) ^{2})}{\text{HeunC}(\alpha_2,\beta_2,\gamma_2,\delta_2,\eta_2,\left( \sin \left( \pi/6  \right)  \right) ^{2})}+ p=0
\label{Wr1}
\end{align}
\end{footnotesize}
\noindent where details on the parameters can be seen in \cite{F1, DF1, DF2, DF3}
\item {Jet modes:} A qualitatively new boundary condition has been used in \cite{DF3} to obtain the so-called primary jet modes. The condition was that of angular singularity which translates into polynomial condition for the solutions of the TAE, i.e.:  
\begin{footnotesize}
\begin{align*}
\footnotesize
 &\frac{\delta}{\alpha}+\frac{\beta+\gamma}{2}+N+1=0\\
 &\Delta_{N+1}(\mu)=0
 \end{align*}
\end{footnotesize}
where $\Delta_{N+1}{(\mu)}$ is tridiagonal determinant \cite{F3}. 
\end{enumerate}
\item{On the TRE:} 
\begin{enumerate}
\item{Black hole boundary conditions}: For any $m$, the solution $R_2$ is valid for frequencies for which $\Re(\omega) \not\in (-\frac{ma}{2Mr_+},0)$ and also that: $\sin(\arg(\omega)\!+\!\arg(r))\!<0$. %(the direction of steepest descent).
\item{Quasi-bound boundary conditions}: For any $m$, the solution $R_1$ is valid for frequencies for which $\Re(\omega) \not\in (-\frac{ma}{2Mr_+},0)$ and also  that: $\sin(\arg(\omega)\!+\!\arg(r))\!>0$.
\end{enumerate}
\end{enumerate}

\subsection{Numerical results}
\begin{figure}[h]
\sidecaption
\includegraphics[scale=.25]{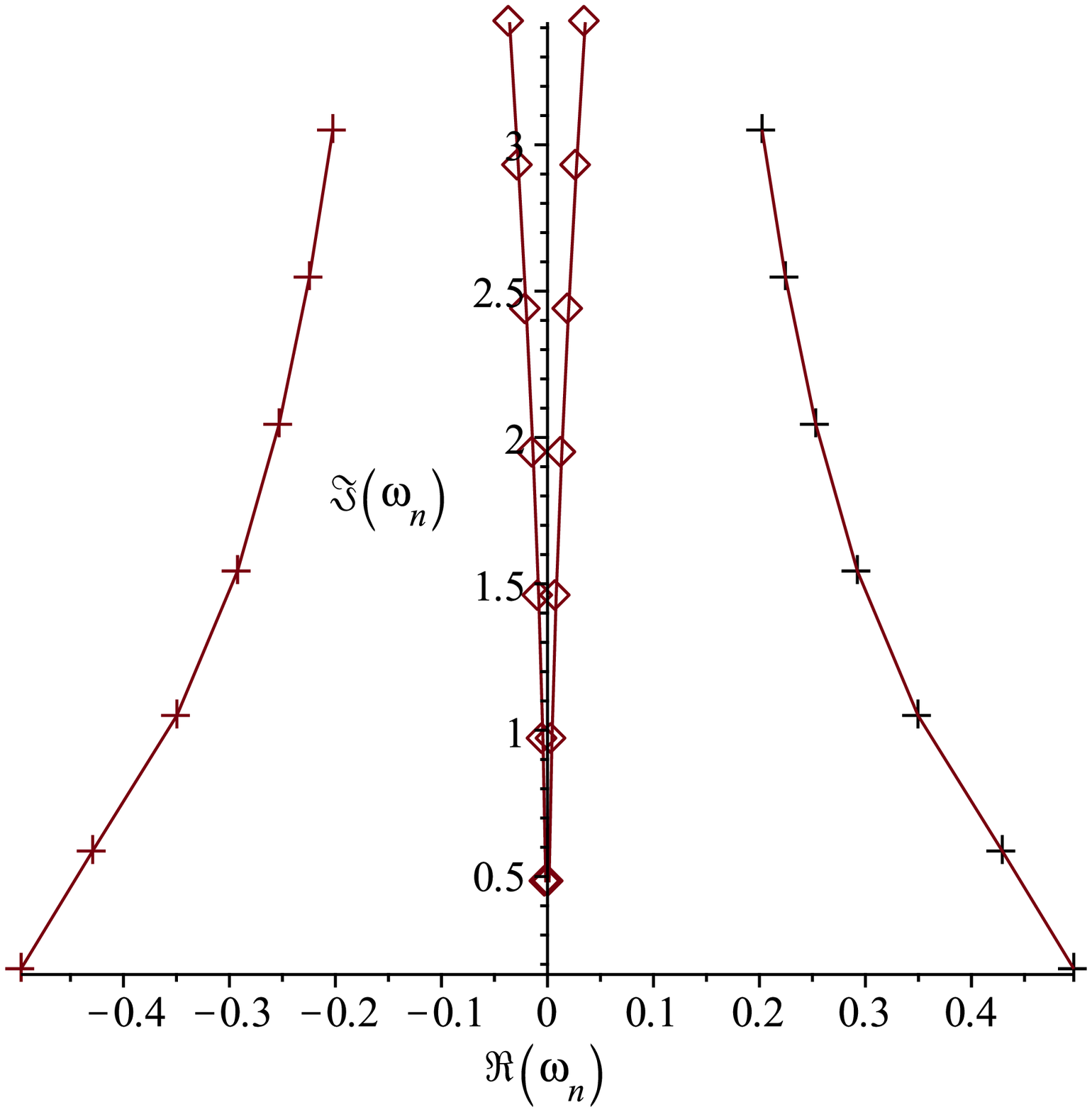}
\includegraphics[scale=.25]{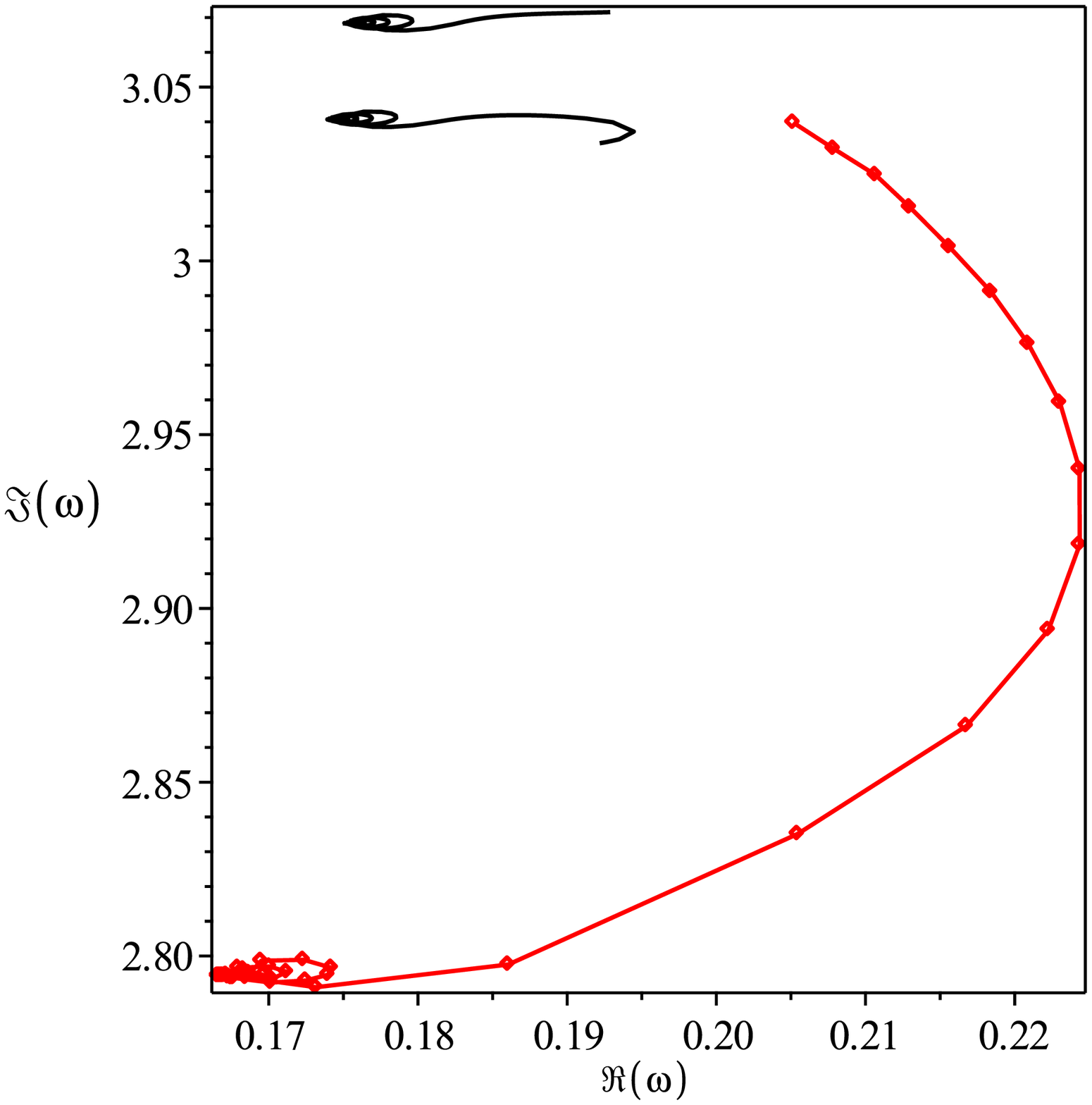}
\caption{Examples of the different spectra obtained from the spectral system. a) Complex plot of the first 7 modes in the QNMs (crosses) and primary jet modes (diamonds) b) QNMs (point-line) and the non-physical spurious modes (solid lines) for $a=[0,M]$}
\label{fig:2}       % Give a unique label
\end{figure}

The so described boundary conditions lead to a two-dimensional spectral system on the unknowns $\omega$ and $E$. Because of the complexity of the confluent Heun functions, we use an algorithm developed by the team to find the roots of the system. The numerical results give different spectra of discrete complex frequencies some of which can be seen on Fig. \ref{fig:2}. As part of our study, we examined how those spectra change with introduction of rotation ($a\ne 0$), up to the limit $a\!\to\! M$, and we tested the numerical stability of the so-obtained frequencies, in order to ensure they represent physical quantities and not a numerical artifact (an example can be seen on Fig. \ref{fig:2} b). )

The physically interesting result are the qualitatively different spectra (Fig. \ref{fig:2} a) ), depending on the boundary conditions imposed on the system, which can be used as an independent tool to discover the nature of the physical object emitting electromagnetic or gravitational waves. 

\section{Conclusion}
In this proceeding we discussed the application of the Heun functions to the problem of quasinormal modes of rotating and non-rotating black holes. We presented some of our latest numerical results, key to which is the development of the theory of the Heun functions and their numerical implementation. 
\bigskip

\begin{acknowledgement}
This article was supported by the Foundation "Theoretical and
Computational Physics and Astrophysics", by the Bulgarian National Scientific Fund
under contracts DO-1-872, DO-1-895, DO-02-136, and Sofia University Scientific Fund, contract 185/26.04.2010, Grants of the Bulgarian Nuclear
Regulatory Agency for 2013, 2014 and 2015.
\end{acknowledgement}

\begin{thebibliography}{99.}%


\bibitem{F1} Fiziev ~P.~P., Class. Quantum Grav. {\bf 23}  2447-2468 (2006)  %, arXiv:0908.4234v4  [gr-qc]

\bibitem{F2} Fiziev ~P.~P., Class. Quantum Grav. {\bf 27}  135001 (2010)%, arXiv:0908.4234v4  [gr-qc]

\bibitem{F3} Fiziev ~P.~P., J. Phys. A.: Math Theor. {\bf 43}  035203 (2010)%, arXiv:0908.4234v4  [gr-qc]

\bibitem{DF2} Fiziev P., Staicova D., Phys. Rev. D \textbf{84}, 127502 (2011) 

\bibitem{HC1}  Heun ~K., Math. Ann. \textbf{33} 161, (1889)

\bibitem{HC2}   Ronveaux, A., ed. Heun's Differential Equations. Oxford, England: Oxford University Press, 1995. 

\bibitem{DF1} %The Spectrum of Electromagnetic Jets from Kerr Black Holes and Naked Singularities in the Teukolsky Perturbation Theory
Staicova D. R., Fiziev P. P. Astrophys Space Sci  \textbf{332}: 385-401 (2011)

\bibitem{DF3} %Numerical stability of the electromagnetic quasinormal and quasibound modes of Kerr black holes
Staicova D. ,  Fiziev P.
Bulgarian Astronomical Journal \textbf{23}, 83, (2015)

\bibitem{DF4}%New results for electromagnetic quasinormal and quasibound modes of Kerr black holes
Staicova D.,  Fiziev P. Astrophysics and Space Science, \textbf{358}:10 (2015)

\bibitem{site} The Heun Project  \textit{http://theheunproject.org/bibliography.html}


% Contribution
%
%\bigskip
%
%
\end{thebibliography}
 \end{document}